\documentclass[runningheads]{llncs}
\usepackage{graphicx}
\graphicspath{ {./images/} }
\usepackage{booktabs}
\begin{document}
\title{Learning Representations for Zero-Shot Retrieval over Structured Data}
%
%
\author{Harsh Kohli\inst{1}\orcidID{0000-0003-1431-6025}}
\authorrunning{H. Kohli}
%
\institute{Compass, Bangalore KA 560066, India\\
\email{harsh.kohli@compass.com}\\}
\maketitle              
\begin{abstract}
Large Scale Question-Answering systems today are widely used in downstream applications such as chatbots and conversational dialogue agents. Typically, such systems consist of an Answer Passage retrieval layer coupled with Machine Comprehension models trained on natural language query-passage pairs. Recent studies have explored Question Answering over structured data sources such as web-tables and relational databases. However, architectures such as Seq2SQL assume the correct table a priori which is input to the model along with the free text question. Our proposed method, analogues to a passage retrieval model in traditional Question-Answering systems, describes an architecture to discern the correct table pertaining to a given query from amongst a large pool of candidate tables.

\end{abstract}
\section{Introduction}

The Stanford Question Answering Dataset (SQuAD) \cite{rajpurkar-etal-2016-squad} has sparked considerable interest within the research community. Ensuing model architectures as well as resources in the form of datasets and challenges have led to remarkable progress in the field of Machine Comprehension. Ultimately, this has led to better Natural Language Understanding systems - including virtual assistants, search engines, and chatbots. The best performing systems consist of large, expressive attention-based deep learning architectures such as BERT \cite{devlin-etal-2019-bert} followed by a specialized classification head or decoder to point to the start and end of an answer span in text \cite{10.5555/2969442.2969540}. In practice, these systems are aided with a distinct IR step in the form of a retrieval based model or ranker which helps select the correct passage or, at least, reduce the search space considerably.

WikiSQL \cite{zhongSeq2SQL2017}, much like SQuAD for passage-based Question Answering (QA), was the pioneering effort for QA from relational databases. The dataset included 80654 hand-annotated questions from 24241 tables across wikipedia. Along with the dataset, the paper introduced a pointer network style encoder-decoder architecture augmented with Reinforcement Learning to convert a natural language question to a SQL query. The model leverages the structure of a typical SQL query as well as information about the corresponding table such as column names to effectively reduce the search space for the generated query. In order to gainfully employ such a system, a preceding step is required to select the pertinent table or shortlist a few contenders where the query is likely to achieve a hit.

We propose an architecture that is able to match the intent of the query with the table in real time. The model leverages information present in each table and its corresponding schema, as well as semantics of the natural language question. In a zero-shot manner, the model is able to generalize to not just unseen questions, but questions from newer tables that are never used in training. The question and table are encoded separately and the vector spaces are aligned in our learning objective. Once these encoders are trained, table embeddings can be indexed and fast vector search libraries such as FAISS \cite{JDH17} can be used to retrieve the candidate tables in real time. We discuss the data used in training the model, as well as experiments conducted and the design choices made in our architecture. Since we use distinct tables in our training, dev, and test data, the idea is not akin to a classification model used to categorize the question as belonging to a specific table. 

\section{Related Work}

Seq2SQL \cite{zhongSeq2SQL2017} explores the problem of Question Answering over relational databases by converting a Natural Language question to a logical form such as an SQL query. Prior to this, studies have investigated QA on other tabular sources such as web tables. TableQA \cite{tableqa} uses a table representation using row-column-value triplets for a table. These triplets are passed along with the question and correct answer to a Memory Network Architecture \cite{10.5555/2969442.2969512}. Neural Enquirer \cite{yin-etal-2016-neural} uses distinct encoders for a query and a table, representations from which are input to a series of executors. Each executor individually models a specific type of operation conditioned on the query. Like Seq2SQL, these approaches presuppose the availability of the correct table. In the practical case of large databases spanning many tables, this is often not true and an intermediate IR step might be required to identify the answer table or atleast reduce the search space. 

The approach described in \cite{10.1145/2872427.2883080} addresses Question Answering across tables. Each row of a table is converted into multiple relational chains and an entity-linking mechanism \cite{Yang2015SMARTNT} is used for candidate chain generation. However, this leads to a very large number of chains especially when multiple tables are involved. These are later pruned via frequency vector based matching by inputting the chain values as well as the question as a query to a search engine (Bing). This step can be prohibitively expensive or infeasible in a lot of scenarios. In our work, we describe a method to retrieve the likely table candidates without relying on an external search engine. 

\section{Dataset \& Preprocessing}

We use WikiSQL \cite{zhongSeq2SQL2017} as our primary data source and also preserve the same train/dev/test split used in training the original Seq2SQL model. However, in order to better align with our learning objectives, we transform the data in the following ways:

\subsection{Train Data}
\label{marker1}

Each question in WikiSQL and the corresponding table is a positive example in our train set. The question is tokenized and converted to lowercase to suit our word embedding model. Column names are extracted and processed similarly, as well as the values in each individual row. This is possible since each table in WikiSQL consists of a small number of rows. For large tables row elements can be sampled and techniques such as stop-words removal can be applied to represent the information in each individual column. Our dataset consists of two different column types - 'text' and 'real'. Real represents tables consisting of a real number in each row. For these tables, an extra token <REAL> is added instead of actual row values and trained as an Out Of Vocabulary (OOV) token. This approach can be extend to different column types in actual databases such as DateTime columns. We use random and hard-negative sampling strategies to sample negatives in our train data.

\subsubsection{Base Word Embedding}

For computing the hardest negatives as well as modelling in later sections, we use the Word2Vec \cite{41224} skip-gram model trained on the English Wikipedia Corpus using hierarchical softmax. The pre-trained model, taken from tensorflow-hub outputs a 500D embedding for each token (word in a sentence). 

\subsubsection{Hard Negative Sampling}
\label{marker2}

For each question in our train corpus, we compute the embeddings from the Word2Vec model. Word embeddings are aggregated using the sqrtn combiner to form a sentence representation for the question. The sqrtn combiner computes the weighted sum of the individual token embeddings and divides by the square root of the sum of squares of the individual weights (set to 1 in our case).

The question embeddings are indexed using the Annoy library for fast Approximate Nearest Neighbor (ANN) search. Post the indexing, for each question in our train set with table id $id1$ (say), the nearest neighbors are retrieved. The closest neighbor $N1$ with a corresponding table id different from that of the original question is treated as the hardest negative. The tuple $(id1, N1, 0)$ is added in our training data with label $0$ indicating that the particular example belongs to the negative class.

\subsubsection{Random Negative Sampling}

For each question in our training set, a random example is selected from amongst the remaining questions. If the new example has a table id distinct from the original question, the tuple consisting of the original table id and the new question is considered a negative example in our training.

\begin{figure}[h]
 \center
  \includegraphics[height=2cm]{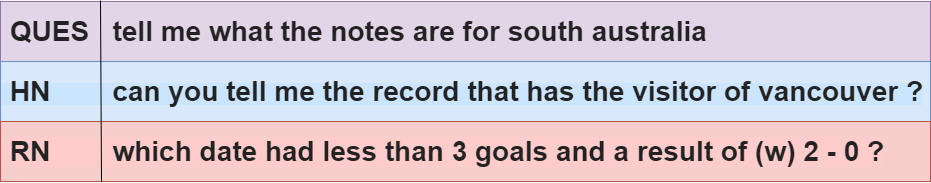}
  \caption{Question, Hard Negative \& Random Negative}
  \label{fig:Examples}
\end{figure}

Figure ~\ref{fig:Examples} shows a natural language question (QUES), a hard negative (HN), and a random negative (RN) for a (question, table) pair in our dataset. Here, the negatives are associated with a table which is different from the question.

\subsection{Dev \& Test Data}

We use the dev and test splits from the WikiSQL dataset. The questions are tokenized and converted to lower case similar to the train set and the corresponding table ids are preserved. The table information such as column names, row values corresponding to each column and column data types are extracted. This information is used to generate representations for each table in our dev and test datasets using our table encoder during and post training. These representations are matched with the question embedding from our question encoders to retrieve a ranked list of tables for each question using ANN search. In our training, the model with the best Mean Reciprocal Rank (MRR) on the dev set is saved and metrics on dev and test sets are reported on the same model.

\begin{figure}
 \center
  \includegraphics[height=8cm]{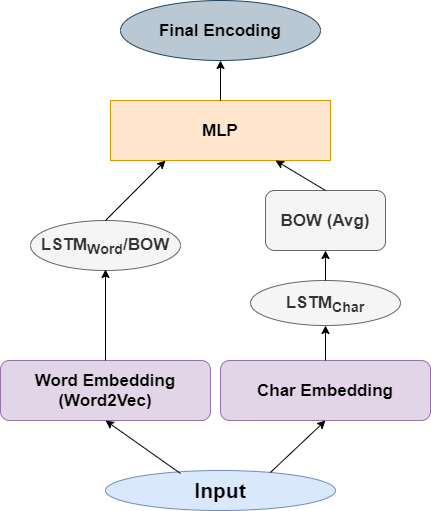}
  \caption{Text Encoder}
  \label{fig:InputEncoder}
\end{figure}

\section{Model Architecture}
\label{marker4}

\begin{figure}
 \center
  \includegraphics[height=6cm, width=\linewidth]{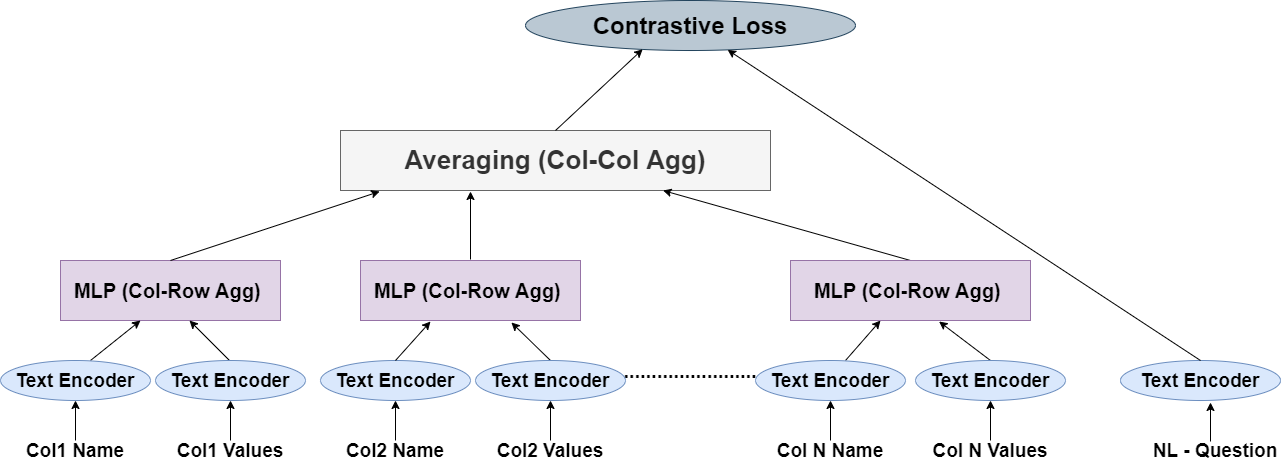}
  \caption{Model Architecture}
  \label{fig:modelDiagram}
\end{figure}

Our model architecture broadly consists of two separate encoders for the tables and the questions. While the Word2Vec model described earlier is used as the base for computing our word embeddings, we augment it with character level embeddings to better represent OOV tokens. The text encoding architecture, which is re-used as a building block for encoding the natural language question, the table column names, as well as the row values is depicted in Figure ~\ref{fig:InputEncoder}. The final training architecture is shown in Figure ~\ref{fig:modelDiagram}.

\subsection{Character Encoder}
\label{marker3}

While popular approaches to character encoding utilize a Char-CNN \cite{10.5555/3016100.3016285} model, we employ a simpler approach consisting of a single LSTM layer \cite{10.1162/neco.1997.9.8.1735}. For input text $T$ to our character encoder consisting of $n$ tokens ${t_1, t_2, ... t_n}$, let us assume a single token in our input $t_i$ consists of $m$ characters $\{c^{i}_{1}, c^{i}_{2}, ... c^{i}_{m}\}$. A 200D embedding layer $E$ is randomly initialized and trained for each individual character.   

\begin{center}
    $E_{x}^{i} = E(c^{i}_{x})$\\
\end{center}

\begin{center}
    $h^{i}(x) = LSTM_{char}(E_{x}^{i}, h^{i}(x - 1))$\\
\end{center}

\begin{center}
    $E_{T} = \frac{1}{n} * \sum_{i=1}^{n} h^{i}$\\
\end{center}

Here, $E_{x}^{i}$ represents the embedding for a character $c^{i}_{x}$ in token $t_i$. The character-encoding LSTM $LSTM_{char}$ is applied to the sequence of character embeddings in the token and $h^{i}(x)$ represents the intermediate hidden state for character $c^{i}_{x}$. The final output of the LSTM $h^{i}$ for each token $t_i$ is computed this way, and embeddings are averaged for each token in text $T$ to get the final representation of our input, $E_{T}$.    

\subsection{Table Encoder}

Table features are extracted as described in Section \ref{marker1}. These include the column names, their corresponding row elements and data types. For real valued columns, instead of the row elements, we train an extra token <REAL> whose embedding is learned during the training process. In this section, we describe the architectural elements used to encode this information and achieve a fixed-length representation for the table.

\subsubsection{Column Name Encoder}

Let $c = \{c_1, c_2, ... c_n\}$ be the column names for a table $T$ with $n$ columns and $m$ rows. Both word and character level embeddings for each snippet in $c$ are computed using the approach described in sections \ref{marker2} and \ref{marker3} respectively. These are combined to obtain the column name embedding $E_c^{i} \in \mathbf{R}^{700}$ for each column in our table:

\begin{center}
    $E_c^{i} = W_c^{i} ; C_c^{i}$\\
\end{center}

Where $W_c^{i} \in \mathbf{R}^{500}$ and $C_c^{i} \in \mathbf{R}^{200}$ are the word and character embeddings and $;$ represents the concatenation operation. 

\subsubsection{Row Values Encoder}

Embeddings for each row entry in the column are obtained similar to the column name embeddings $E_r^{ij} \in \mathbf{R}^{700}$ (where $i$ is the column number and $j$ represents the row number). The embeddings for the individual rows are combined using a bag-of-words style averaging.

\begin{center}
    $E_r^{i} = \frac{1}{m} * \sum_{j=1}^{m} E_r^{ij}$\\
\end{center}

\subsubsection{Row-Column Aggregation}

The column name embedding $E_c^{i}$ and row values embedding $E_r^{i}$ are concatenated and passed through a single dense layer to obtain an intermediate column representation $e_i \in \mathbf{R}^{1000}$. In our ablation experiments where the character embeddings are not considered, $E_c^{i}$ and $E_r^{i}$ are directly concatenated to get $e_i$. $e_i$ is then passed through a multi-layer perceptron (MLP) to obtain the final column embedding $E_i \in \mathbf{R}^{500}$.

\begin{center}
    $E_i = MLP(e_i)$\\
\end{center}

The MLP consists of 2 hidden layers of sizes 750 and 500.

\subsubsection{Column-Column Aggregation}

Finally, the set of column embeddings $E=\{E_1, E_2... E_n\}$ is averaged to get the final table embedding $E_T \in \mathbf{R}^{500}$.

\begin{center}
    $E_T = \frac{1}{n} * \sum_{i=1}^{i} E_i$\\
\end{center}

Throughout this process of obtaining the final table representation, we use averaging for aggregation and dense/MLP layers to add expressiveness to our model whenever required. We avoid using a sequence model such as LSTM (with the exception of the character embedding module) as the relative ordering of the rows in a column, and the ordering of the columns themselves are not important.

\subsection{Question Encoder}

For obtaining the question representation, we directly use the text encoding method described in Figure ~\ref{fig:InputEncoder}. We get the word and character embeddings for each token in the question as described previously. Character-level representation for the question is obtained by averaging the $LSTM_{char}$ outputs for each word, whereas for the word-level representation we adopt two strategies - BOW (averaging the word embeddings) and $LSTM_{word}$. In the bag-of-words model, a 2-layer MLP (each layer of size 500) is used to get the final word-level representation after averaging. Alternatively, $LSTM_{word}$ is an LSTM layer of size 500 used to encode the sequence of token embeddings from Word2Vec. Either way, we obtain a 500D representation for the word encoding and combine that with the 200D character-level representation of the question. A single dense layer is applied on the combined 700D embedding to get the final question representation $E_Q \in \mathbf{R}^{500}$. Like in our table encoder, this final dense layer is skipped in experiments where the character embeddings are not considered.

\section{Optimization \& Model Parameters}

For each tuple in our training set consisting of a question, table, and label, we compute the table embedding $E_T$ and question embedding $E_Q$ using the approach described in Section \ref{marker4}. We use a contrastive loss \cite{1640964} to tune our network which selectively optimizes for negative examples when they are within a certain distance or margin of one another. The loss is defined as follows:

\bigskip

\begin{center}
loss = $\frac{1}{2} (1 - y) (d(E_T, E_Q))^2 + \frac{1}{2}y(max\{0, m-(d(E_T, E_Q))^2\})$
\end{center}

\bigskip

Here, $y$ is the label whereas $d(E_T, E_Q)$ represents the euclidean distance between embeddings $E_T$ and $E_Q$. The margin $m$ is set to 0.5 and the inputs are L2-Normalized at the encoding step

The Adam optimizer \cite{Kingma2015AdamAM} is used to train our network. The learning rate is set to $10^{-4}$ and a batch size of 128 is used in our training. Wherever applicable, such as the various MLP or LSTM layers, the rectified linear unit (ReLU) is used as our activation function.

\section{BM25 Baseline}

The classic and effective BM25 ranking function is used as a baseline to our approach. All tokens corresponding to a table including individual row values and column names are extracted and indexed using the Gensim package \cite{rehurek2011gensim}. Question words are tokenized similarly and top matching tables are retrieved using the BM25 scoring function.

\begin{table*}[htbp]
\centering
\begin{tabular}{|c|c|c|c|c|c|c|}
\hline \bf Model & \bf MRR & \bf P@1 & \bf P@5 & \bf P@10 & \bf P@50 & \bf P@100\\ \hline
BM25 & 12.50 & 9.11 & 14.96 & 18.74 & 29.47 & 35.38\\
Hardest\_BOW & 9.31 & 4.29 & 12.27 & 18.68 & 43.60 & 56.86\\
Random\_BOW & 19.50 & \bf 10.20 & 27.31 & 38.61 & \bf 71.40 & 84.39\\
H+R\_BOW & \bf 19.75 & 10.16 & \bf 28.22 & \bf 39.18 & 71.15 & \bf 85.17\\
H+R\_BOW\_char & 14.91 & 6.58 & 21.14 & 31.61 & 64.97 & 79.85\\
H+R\_LSTM & 16.80 & 8.12 & 23.37 & 35.00 & 69.59 & 84.12\\
\bottomrule
\end{tabular}
\caption{Dev Results (2716 Tables)}
\label{devresults}
\end{table*}

\begin{table*}[htbp]
\centering
\begin{tabular}{|c|c|c|c|c|c|c|}
\hline \bf Model & \bf MRR & \bf P@1 & \bf P@5 & \bf P@10 & \bf P@50 & \bf P@100\\ \hline
BM25 & 11.29 & \bf 8.46 & 13.87 & 16.34 & 24.88 & 29.92\\
Hardest\_BOW & 6.10 & 2.37 & 8.18 & 12.48 & 31.02 & 43.51\\
Random\_BOW & 14.06 & 6.90 & 19.17 & 27.92 & 57.39 & \bf 72.06\\
H+R\_BOW & \bf 14.24 & 6.95 & \bf 19.78 & \bf 28.65 & \bf 57.66 & 72.00\\
H+R\_BOW\_char & 9.62 & 3.68 & 13.14 & 21.00 & 50.34 & 66.17\\
H+R\_LSTM & 11.52 & 5.00 & 15.78 & 24.44 & 54.48 & 70.53\\
\bottomrule
\end{tabular}
\caption{Test Results (5230 Tables)}
\label{testresults}
\end{table*}

\section{Experiments \& Results}

We experiment with the various architectural design choices as well as negative sampling techniques to empirically determine the best configuration. Random\_BOW and Hardest\_BOW denote models where only random and hard negatives are used in training respectively along with the Bag-of-Words question encoder. Similarly, H+R\_* are models where both random and hard negatives are used. *\_char denote models where character embeddings are used alongside the word embeddings from Word2Vec and *\_lstm represent iterations where the LSTM model is used for our question encoder instead of BOW.

The dev and test results from our training runs are summarized in Tables \ref{devresults} and \ref{testresults} respectively. We report the mean reciprocal rank (MRR) as well as precision at top K (P@K). The annoy library is used for indexing and searching on the dev and test tables. The disparity in metrics between the dev and test set is because of the larger number of tables in our test set (5230 vs 2716) leading to a larger search space.

We note that while most models outperform the BM25 baseline at nearly all precision levels (K), the difference is more pronounced at higher values of K. This is because, for the more obvious questions comprising a direct token match, BM25 does a reasonable job of identifying the correct table. However, for the harder questions, the encoding architecture built atop a pre-trained word embedding layer is able to learn a better representation thus enabling a more effective ranking. Recall for tables in these cases is more likely in our approach, albeit at larger K values.

The bag-of-words model with both hard and random negatives achieves the best results. The inclusion of character embeddings does not improve performance, possibly due to a small fraction of OOV words especially in the questions and column names. Unlike the word embeddings from Word2Vec, our character embeddings were not pretrained on a large corpus and randomly initialized. We hypothesize that due to limited training data (56k unique questions), our model does not succeed in learning a  good representation which adversely affects performance. The reasoning also applies to our more expressive LSTM encoder experiments as they do not perform as well on our metrics as the simple bag of words with a feed-forward network. However, this might change if we were to train on a larger corpus or have a greater number of rare words in our test sets. Including hard negatives improves performance over the model trained on only random negative samples, however a model trained on hard negatives alone performs poorly. 

\section{Conclusion \& Future Work}

In this paper, we describe a method to encode a database table and a natural language question and align their vector spaces. Empirically, we determine the ideal combination of input data and architectural blocks for our particular dataset. Our best model is able to reduce the search space to a 100 tables with over 85\% precision on the dev set and 72\% precision on the test set. This could potentially reduce overhead on a downstream Seq2SQL model as it needs to query fewer tables to retrieve the desired answer.

We observe from our experiments that simpler, less expressive models tend to perform better which indicates that our model could benefit from more training data. Incorporating embeddings from more recent pretrained architectures such as BERT might help improve performance further. However, this might prove expensive for an L1 retrieval model, if we were to encode large tables with many columns. Various other sampling and training strategies, such as semi-hard negative sampling coupled with a triplet loss objective \cite{1640964} can be used to tune our model further and achieve better results for our task.

Our model performs a zero-shot generalization on test data. However, including some in-domain data in a distinct fine-tuning step would likely improve results. In most production systems the tables and their schemas are fairly static and the model would rarely have to generalize to an unknown table, thus making this a one-time annotation effort. Studies on schema driven dialogue systems \cite{DBLP:conf/aaai/RastogiZSGK20} have shown that including natural language descriptions of schema elements such as slots, values, intents, service descriptions etc., and incorporating them in the training objective helps the models better perform zero-shot or few-shot generalizations to unknown schemas. Similarly, table descriptions along with descriptions of each column can be expected to boost model performance on unseen tables.

%
%
%
\bibliographystyle{splncs04}
\bibliography{mybib}
%

\end{document}